\documentclass[draftcls,11pt,onecolumn,a4]{IEEEtran}
\newlength{\intwidth}

\newcommand{\rel}{\mbox{Rel}}

\newcommand{\E}{\ensuremath{\mathrm{E}}}
\newcommand{\Ex}{\ensuremath{\mathrm{E}}}
\newcommand{\var}{\ensuremath{\mathrm{var}}}
\newcommand{\cov}{\ensuremath{\mathrm{cov}}}
\newcommand{\Cov}{\ensuremath{\mathrm{Cov}}}

\newcommand{\Var}{{\mathrm{Var}}}

\usepackage{mathrsfs}
\usepackage{graphics,graphicx,bm,amsmath,amssymb}
\usepackage{pst-all,amstext,amsfonts,epsf,epsfig,pifont}
\begin{document}
%
\title{Modelling and Estimation of Covariance of Replicated Modulated Cyclical Time Series}
%
%
\author{Sofia~C.~Olhede and Hernando Ombao 
\thanks{This work was supported by an EPSRC grant EP/I005250/1 and
US NSF DMS and NSF SES grants.}
\thanks{S. Olhede is with University College London,
Department of Statistical Science, Gower Street,
WC1E6BT, London, UK (s.olhede@ucl.ac.uk). Tel:
+44 (0) 20 76798321, Fax: +44 (0) 20 73834703. H. Ombao is with University of California at Irvine,
Department of Statistics, Irvine, CA 92618, USA (hombao@uci.edu). Tel: +1 (949) 824-5679.}}

\markboth{Statistical Science Research Report 316}{Olhede \& Ombao: Modeling Replicated Modulated
Cyclical Time Series}
\maketitle

\begin{abstract}
This paper introduces the novel class of modulated cyclostationary processes, a class of
non-stationary processes exhibiting frequency coupling, and proposes a method of their estimation from repeated trials. Cyclostationary processes also exhibit frequency correlation but have Lo\`eve spectra whose
support lies only on parallel lines in the dual-frequency plane. Such extremely sparse structure
does not adequately represent many biological processes. Thus, we propose a model that, in the time
domain, modulates the covariance of cyclostationary processes and consequently broadens their frequency support  in the dual-frequency plane.
The spectra and the cross-coherence of the proposed modulated cyclostationary process are first
estimated using multitaper methods. A shrinkage procedure is then applied to each trial-specific
estimate to reduce the estimation risk.

Multiple trials of each series are observed. When combining information
across trials, we carefully take into account the bias that may be introduced
by phase misalignment and the fact that the Lo\`eve spectra
and cross-coherence across replicates may only be ``similar" -- but not necessarily identical --
across replicates.
The application of the inference methods developed for the modulated cyclostationary model to EEG data also demonstrates
that the proposed model captures statistically significant cross-frequency interactions, that ought to be further examined by neuroscientists.

\end{abstract}
\vspace{0.1in}

\section{Introduction \label{intro}}
The primary contribution of this paper is a novel model for nonstationary time series
that rigorously explains a particular dependence structure in the spectral domain that
current models cannot handle.  This model is inspired by an electroencephalography (EEG)
data set that exhibits coupling between frequency bands. We develop a time-domain representation of the model and derive the corresponding Lo\`eve spectrum
that captures the desired features.  Methods for estimating such features are proposed, and challenges inherent in the estimation discussed.

Nonstationary processes are ubiquitous in practical applications, we mention for example vibratory
signals \cite{Kim2005,Zhang2003}, free-drifting oceanic instruments \cite{Lilly2012}, and
various neuroscientific applications \cite{Cranstoun,Ombao2001}. Under stationarity, there are
several procedures available for obtaining mean-squared consistent estimates of the spectrum.
When the assumption of stationarity is rendered invalid, this implies that the second-order structure
of the process is usually more complicated. For example, in the case of harmonizable processes,
the spectrum is now defined on a {\em dual-frequency} plane $(-\frac{1}{2},\frac{1}{2}) \times (-\frac{1}{2},\frac{1}{2})$ -- in contrast to just the interval $(-\frac{1}{2},\frac{1}{2})$ used in the stationary case. Thus, alternative assumptions need to be made
to enable estimation, because it is generally not possible to estimate the Lo\`eve
spectrum at the ${\cal O}(T^2)$ pairs of the fundamental Fourier frequencies from time series of
length $T$ without imposing additional constraints.

Many models for non-stationary processes have been developed. In classical signal processing,
the evolutionary behavior of the spectrum is modeled assuming substantial smoothness in the form
of the time-variation of the process, see e.g., \cite{Dahlhaus2000,Silverman1957}. These models
form the backbone of using the popular windowed Fourier analysis which implicitly assumes that the signal
is approximately stationary within the time window of analysis and that the spectrum of the time series is also smooth within the same window. If the smoothness assumption is not suitable, then replicates of the series are necessary
to enable estimation. Unfortunately, in order for replication to be successfully utilized in the
estimation, we must have time sample replicates that exhibit both perfect time alignment
and exactly the same form of nonstationarity across replicates.

In this paper, we propose an analysis framework that is valid for replicated time series (which is fairly
typical in designed experiments) where we repeatedly observe phenomena that are ``similar" across
replicates but {\em not} necessarily identical, and the model must include various realistic features of the replication. First, there may be phase-shifts between the replicates
which can complicate the estimation of the Lo\`eve spectrum (or the coherency).
Second, the dependence between a fixed pair of frequencies (or frequency bands) may not perfectly
replicate at each trial. Hence, our model must account for this variation and our estimation methods
must have the ability to recover this variation. In Section \ref{2ndorder}, we shall discuss second
order modeling of a time series and then focus on frequency correlation descriptions of the signal
in Section \ref{dual-freq}.

In order to fully understand the underlying stochastic process that produces the observed
time series, it is important to develop time domain models that yield the features or shapes
that we observe in the frequency domain.
In analyzing our electronencephalogram (EEG) data, we observe spectral domain structures
that are in parts reminiscent of existing models -- e.g. cyclostationary
models \cite{Gladyshev1961,Gardner2006,Gardner} -- e.g. that have a baseline frequency of importance and where
harmonics with this period may correlate. However, the structures that we observe do {\em not} exactly
fit the current models because they show a significant spread in the dual-frequency plane. Thus,
motivated by the results from a preliminary analysis of the EEGs, we introduce a time domain class
of processes that may reproduce such structure. In Section \ref{mod-cyc}, we develop the proposed
{\em modulated} cyclostationary process and discuss its  relationship to modulated oscillations,
see e.g. Voelcker \cite{Voelcker1966a}. Properties of the replication are explicitly discussed in
Section \ref{rep-mod}.

Guided by the features of spread in dual-frequency and the variation and
similarity across replicated trials, we develop an appropriate algorithm for estimation in Section \ref{Est}.
First, as suggested by the data, we assume some smoothness of the covariance of the process
in frequency to form a replicate-specific estimate of the Lo\`eve spectrum.
We shall also examine the validity of this assumption in Section \ref{MT-Est}.
One possibility would be to compute the average spectrum across the replicates after denoising (testing for non-zero contribution) and then perform other linear analysis steps across replicates.
These steps are described and discussed in Section \ref{MT-SVD}.

It is critical to understand the family of Lo\`eve coherence functions (or matrices sampled at a set of frequencies and collected in a vector).
To explain and recover the possible structures present in these matrices, we could perform a singular
value decomposition directly on the estimated matrices derived from each replicate.
However, it is not a good idea to apply simple linear methods to the raw Lo\`eve spectra because
of phase shifts between replicates that will act antagonistically, as is discussed explicitly in
Section \ref{TMT-SVD}. Instead we threshold each replicate-specific estimated Lo\`eve spectrum
on its magnitude. We take account of the Hermitian symmetry of the spectrum as redundant tests
makes the testing procedure deteriorate and additionally remove the frequencies that are known
to contain unimportant neuroscientific information.
Very few of the singular vectors are important to the family of matrices and we can recognize a family
of possible structures. This provides a satisfactory description of the variability in frequency
across the replicates. Finally, to arrive at a good understanding of a ``population" of "similar"
processes across replicates (rather than a trial-specific process), we must carefully consider the
sequential ordering of the replicates. As the true underlying
Lo\`eve coherence might have evolved over the course of the experiment, we anticipate
the estimated Lo\`eve coherence to also follow a similar behavior. Moreover, it is reasonable to
assume that the true Lo\`eve coherence is slowly changing throughout the duration of the
experiment and thus we anticipate the Lo\`eve coherence to be similar between ``neighboring"
replicates.
We discuss this structure and use clustering methods to determine its form.

We conclude by analyzing neuroscientific data with the characteristic of the models
developed in this paper.
Our choice of EEG data example motivated our development of the model-class of modulated
cyclostationary process.
EEGs have been important tools in clinical research to investigate underlying brain processes.
This particular visual motor EEG data set has been analyzed by others using classical
spectral methods: e.g., Freyermuth et al. (2010) developed a wavelets mixed effects model to estimate
both the condition-specific spectra and variation in the brain responses across trials; and
Fiecas et al. (2011) explored the brain effective connectivity structure in a network of 12 channels
using generalized shrinkage methods for estimating classical partial coherence. Here, we demonstrate that the
previous analyses missed very interesting oscillatory features -- specifically the coupling between
the alpha and beta oscillations -- that we believe ought to be investigated by neuroscientists.
We conclude by discussing how to encompass additional observed features
in the model and put forth possible physical interpretations of the observed phenomena.

\section{Second-Order Modelling}\label{2ndorder}
We shall discuss the analysis of the covariance of a zero-mean time series $X_t$. We shall therefore model the dual-time autocovariance function \cite{Hindberg2007} of $X_t$ given by
\begin{equation}
s(t,l)=\Cov\left[X_t, X_{t-l}\right]=\E\left[X_t X_{t-l}\right].
\end{equation}
If $X_t$ is second-order stationary  then $s(t,l)$ takes the simpler form $\widetilde{s}(|l|)$.  It is convenient \cite{Sanei} to represent $X_t$ in the frequency domain and one can write
down the Cram\'er representation to be
\begin{equation}
\label{Cramer}
X_t=\int_{-\frac{1}{2}}^{\frac{1}{2}}\exp(i2\pi ft)dZ(f),
\end{equation}
where $\{dZ(f)\}$ is a zero-mean orthogonal increments process with variance
$\Var\{dZ(f)\}=S(f,f)\,df$ if the process has a smooth spectrum. If we define the Fourier transform of $\widetilde{s}(|l|)$
to be $\widetilde{S}(f)$, then for stationary processes $S(f,f)\equiv \widetilde{S}(f)$.
The total ``energy'' or variance of $X_t$ admits a representation of integrating $S(f,f)$ over all $f$. Therefore, if a given frequency $f$ has a relatively large value of $S(f,f)$
associated with it, then we
say that oscillations of frequency $f$ dominate the signal. Furthermore
the autocovariance sequence of a
stationary process is a Fourier transform pair with the spectrum.
Therefore the spectrum fully represents a Gaussian time series because series
are fully specified by their
autocovariance sequence. While useful in a general setting and the focus
of several studies, the stationarity
assumption is not valid for many forms of data. We now outline how the representation
of a stationary covariance can be generalized to a larger class of
processes that can capture realistic features.

\subsection{Dual-frequency representation of Covariance}\label{dual-freq}
As noted, EEG signals are characterized in terms of their frequency content.
While the spectrum already tells us a great deal about the presence of oscillatory patterns
and behavior, it does not capture other interesting and more complex oscillatory properties
of these signals, such as the dependence of the coefficients across the Fourier frequencies.
In general, for non-stationary signals, such correlations are present, even if impossible for stationary signals and
are important to characterize.
It is therefore necessary to define a measure of the degree of correlation between the increment process at two different frequencies.
 For this purpose, we use the Lo\`eve coherence (a measure of the covariance) using the formalism of harmonizable processes \cite{Loeve}.
A time series $X_t$ belongs
to the harmonizable class if it has the Cram\'er-Lo\`eve representation
\begin{eqnarray} \label{Eq:HarmonRep}
X_t = \int_{-\frac{1}{2}}^{\frac{1}{2}}\exp (2\pi i f t) dZ(f),
\end{eqnarray}
where the increment random process $dZ(f)$ still has zero-mean but also satisfies
\[
\Cov[dZ (f_1), dZ(f_2)] = \Ex [dZ (f_1 ) dZ^{\ast} (f_2) ]= S(f_1,f_2) df_1
df_2 ,  \label{loevspect}
\]
where $dZ^{\ast}(f)$ denotes the complex-conjugate of $dZ(f)$ and
$S(f_1,f_2)$ is a complex-valued scalar quantity called the Lo\`eve
spectrum \cite{Hindberg2007}. 
It may appear that Equation \eqref{Cramer} strongly resembles Equation \eqref{Eq:HarmonRep}
in form, but the introduction of the correlations between frequencies significantly
modifies the appearance of realizations $\{X_t\}$.
%
The Lo\`eve spectrum easily relates to the primary quantity
of the auto-covariance of $X_t$ from the following relationship
\begin{equation}
\label{decomp_energy}
s(t,l)=\cov[X_t, X_{t-l}]=\int_{-\frac{1}{2}}^{\frac{1}{2}} \int_{-\frac{1}{2}}^{\frac{1}{2}} S(f_1,f_2)e^{2i\pi
[(f_1-f_2)t+f_2l]}\,df_1\,df_2.
\end{equation}
Thus, the covariance of $X_t$ with $X_{t-l}$ across time $t$ is still decomposed
to contributions across frequency. However, in general, we also need to consider the
cross-dependencies {\em between} different frequencies.

The Lo\`eve coherency at the pair of frequencies $(f_1, f_2)$, denoted
$\tau(f_1,f_2)$,
is defined to be
\begin{eqnarray}\label{Eq:cohcrosscoh}
\tau(f_1,f_2) = \frac{S(f_1,f_2)}{\sqrt{S(f_1,f_1) S(f_2,f_2)}}=\rho^{1/2}(f_1,f_2)e^{-i\phi(f_1,f_2)},
\end{eqnarray}
where the Lo\`eve coherence at $(f_1, f_2)$ is $\rho(f_1,f_2) = \left
| \tau(f_1,f_2) \right |^2$, and the Lo\`eve coherency phase is $\phi(f_1,f_2)$.
The Lo\`eve coherence values lie in the range $[0,1]$ where values
closer to $1$ indicate a stronger
linear dependence between the random coefficients $ dZ(f_1) $
and $ dZ(f_2)$. It can also be interpreted as
the proportion variance of $ dZ(f_1)
$ that is explained by a linear relationship to $ dZ(f_2) $, e.g.
\begin{equation}
\label{eq:linear}
dZ(f_1)=\tau dZ(f_2)+\varepsilon(f_1,f_2).
\end{equation}
where $\varepsilon(f_1,f_2)$ is uncorrelated with $dZ(f_2)$.
When the Lo\`eve coherence is close to $1$, one interpretation is that increased
oscillatory activity at the
$f_1$ frequency band is strongly associated with an excitation (or an inhibition) of
oscillatory activity at the $f_2$ band. When $|\tau|<1$ then the interpretation of Equation \eqref{eq:linear}
is that the variance of $dZ(f_1)$ is ``partially explained'' by $dZ(f_2)$.
Some care must be applied, because the complete relationship between two complex-valued quantities may be {\em widely-linear} rather than linear, see. e.g. \cite[Eqn. 12]{Hindberg2007}, and the full set of relationships takes the (widely linear) form of
\begin{equation}
dZ(f_1)=\tau dZ(f_2)+\zeta dZ^\ast(f_2)+\varepsilon(f_1,f_2),
\end{equation}
where  $\varepsilon(f_1,f_2)$ is both uncorrelated to and has zero relation with $dZ(f_2)$.
Because $dZ^\ast(f_2)\equiv dZ(-f_2)$ we can take this extra information into account by considering the correlation between the full set of frequencies (e.g. also positive frequencies with negative frequencies) when analyzing $dZ(f)$, rather than calculating the Lo\`eve coherence of $dZ(f)$ and $dZ^\ast(f')$ and the complementary Lo\`eve coherence over a restricted set of frequencies\footnote{Please see \cite{Hindberg2007} for a more complete discussion of the generalized spectral coherence.}.

\begin{figure}[t]
  \centering
   \begin{minipage}[]{0.30\textwidth}
      \centering
       (a) \\
       \includegraphics[scale=0.7]{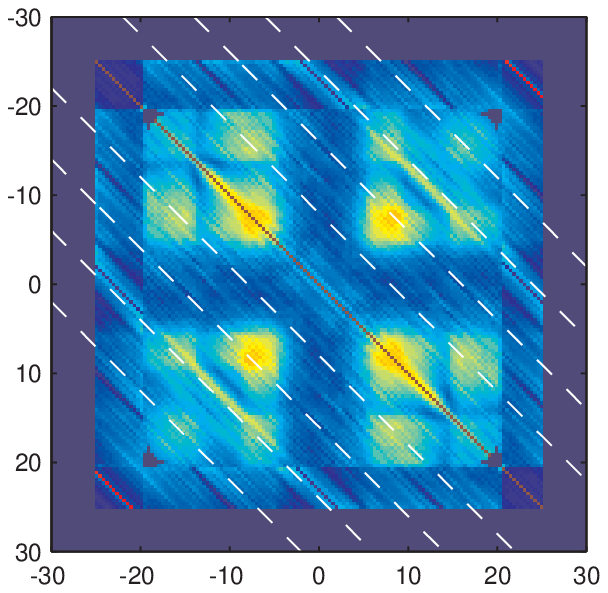}
       \end{minipage}
 \begin{minipage}[]{0.30\textwidth}
      \centering
       (b) \\
       \includegraphics[scale=0.7]{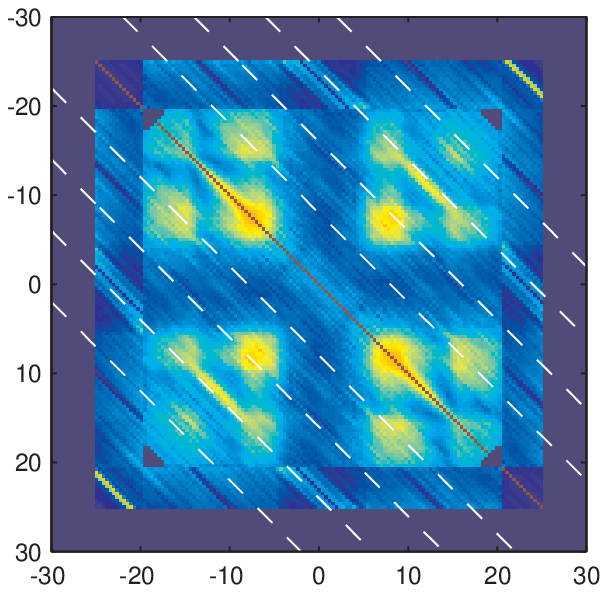}
       \end{minipage}
       \begin{minipage}[]{0.30\textwidth}
      \centering
       (c) \\
       \includegraphics[scale=0.7]{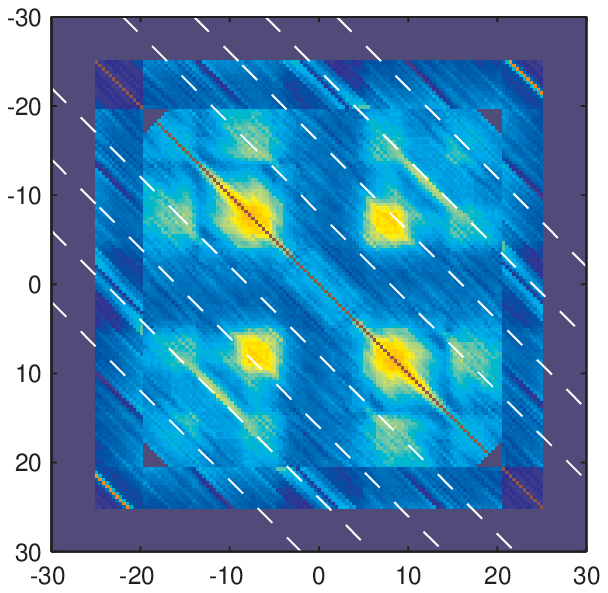}
       \end{minipage}
   \caption[1]{Each plot is the square root of the Lo\`eve coherence averaged across separate blocks of 20 trials. The $x$ and $y$ axes are for frequencies in the range $(-30, 30)$ Hertz.
   Plot (a) corresponds to trials 1--20; Plot (b) corresponds to trials 41--60;
Plot (c) corresponds
    to trials 81--100. These patterns are highly replicable across the aggregate trials.}
  \label{fig3}
\end{figure}

Having extended the classical spectrum to the Lo\`eve spectrum (with particular characteristic features)
we will now need to develop inference methods for this scenario.
There are two natural sets of assumptions one can make concerning the Lo\`eve spectrum.
First, one can assume $S(f_1,f_2)$ to be a smooth function of $f_1$ and $f_2$ and adapt statistical
procedures from non-parametric function estimation. Second, one can impose some form of sparsity on the
support of $S(f_1,f_2)$ -- e.g., assume that the support of $S(f_1,f_2)$ is {\em only} a set of lines in
the dual-frequency domain \cite{Lii1998,Lii2002} which is the case if the process is {\em cyclostationary}
\cite{Gardner2006,Gardner}.
Of course, our model assumptions should be as realistic as possible and must closely match
the data being analyzed. Our motivation for these methods are features encountered in neuroscience, but we envision their application to be more generic to processes exhibiting strong harmonics such as (bio-)acoustic signals.
In Figure \ref{fig3}, we see these features where the raw non-parametric estimates of the Lo\`eve spectrum with limited smoothing
have periodic components that display time-variation and some ``broadening'' which is a deviation from crisply-defined
spectral lines obtained from the class of cyclostationary processes.
Some care should be taken when interpreting these figures, as when averaging over trials we are getting ``sample characteristics'' over the full mixture existing in the population of matrices over the population, a topic we shall revisit.

\subsection{Brief Review of Cyclostationary Processes}\label{cyc}
We first discuss the model structures that will allow us to describe correlations
in frequency like in Figure \ref{fig3}. Firstly, there is structure consistent
with support on lines parallel to the diagonal. The simplest such process
is a cyclostationary process (see e.g. \cite{Gardner2006}). The covariance of  cyclostationary process repeats perfectly in time over a cyclical
period of $D>0$, say.
In fact the Lo\`eve spectrum of a cyclostationary process takes the form (see e.g. \cite{Gladyshev1961,Gladyshev1963})
for some integer value $C\ge 1$
\begin{equation}
\label{cyclostationary}
S(f_1,f_2)=\sum_{c=-C}^C S_c(f_2) \delta\left(f_1-f_2-\frac{c}{D}\right),
\end{equation}
where $\{S_c(f)\}$ are $2C+1$ functions defined for $f\in [-1/2,1/2]$, with respective inverse Fourier transforms $s_c(l)$.
The sum in \eqref{cyclostationary} must be symmetric in $c$ as the observed
process $X_t$ is real-valued.
This Lo\`eve spectrum is equivalent to a time domain covariance of
\begin{eqnarray}
s(t,l)=\int_{-\frac{1}{2}}^{\frac{1}{2}} \int_{-\frac{1}{2}-f}^{\frac{1}{2}-f} S(f+\nu,f)e^{2i\pi (\nu t+f l)}\,d\nu \,df
=\int_{-\frac{1}{2}}^{\frac{1}{2}}\sum_{c=-C}^C S_c(f) e^{2i\pi (\frac{c}{D} t+f l)}\, df=\sum_{c=-C}^C
s_c(l)e^{2i\pi \frac{c}{D} t}, 
\label{cyclostationary_time}
\end{eqnarray}
where $\varsigma_c(t,l)=s_c(l)e^{2i\pi \frac{c}{D} t}$ and
each contribution to the covariance $\{\varsigma_c(t,l)\}$ has associated
period $D/c$, and where each contribution represents a line in the Lo\`eve
spectrum, spaced at $c/D$ from the diagonal.
%

To understand how line components aggregate in the representation, consider the example
\begin{eqnarray}
\label{adding}
X_t&=&\epsilon_t+a(f')\cos(2\pi f' t+\phi_0(f'))\epsilon_t+a(2f')\cos(2\pi
2f' t+\phi_0(2f'))\epsilon_t
\end{eqnarray}
which is a twice-modulated stationary process $\epsilon_t$. Then, the Lo\`eve spectrum of $X_t$ has contributions at $f_1=f_2$ and $f_1-f_2=\pm cf'$.
If the spectrum of $\epsilon_t$, denoted $S_\epsilon(f,f)$, exhibits a peak at $f',$ then this peak generates
a whole structure from the modulation of several peaks. See Figure \ref{fig1}(a),
which illustrates how one peak in the stationary process is propagated onto
several points by the modulation, where the ${\mathcal{S}}_c(f_1,f')$ are
determined by the spectrum of $\epsilon_t$ and the amplitudes $a(f')$  and
phases $\phi(f')$ and so on. The linear mechanism of Equation \eqref{adding} mimics
the behaviour of non-linear systems of coupled oscillations, see for example
the full discussion in \cite[pp.~25]{Priestley1988}. However, while the cyclostationary
model allows us to understand the mechanisms of the second order structure
of non-linear models with coupled oscillations, and describes the ``skeleton''
of the features present in the EEG data, it is not sufficient to
describe the covariance properties of the observed data in more detail, as
evidenced from Figure \ref{fig4}(a) and (c) showing individual trial estimated
Lo\`eve coherences.

\subsection{Modulated Cyclostationary Processes}\label{mod-cyc}
We now develop the novel class of modulated cyclostationary processes. In doing so,
we wish to retain the assumption of simplicity of the Lo\`eve spectrum but
do not constrain the true geometry of the spectrum to live on the exact lines as in
cyclostationary process. Hence, we relax Equation \eqref{cyclostationary_time} to
\begin{eqnarray}
\label{cyclostationary2}
s(t,l)&=&\sum_{c=-C}^C\varsigma_c(t,l)=\sum_{c=-C}^C a_c(t)s_c(l)e^{2i\pi
t c/D}.
\end{eqnarray}
This covariance is nearly identical to Equation \eqref{cyclostationary_time}
but now each line contribution in the covariance has been modulated by the
time-varying amplitude $a_c(t)$.

This is very similar to the notion of uniformly modulated processes; see
e.g. \cite{Priestley1988} but acting on the components of the covariance sequence of
the process, each component being shifted in frequency by $c/D$.
Therefore, unlike uniformly modulated processes, here we allow each
extra line to be modulated {\em differently}. The combination of these actions is
a generalization of the uniformly modulated processes.
The reason for this is two-fold. First, we are modulating each spectral
correlation component $\{s_c(l)\}$ using a {\em different} function, and
so our process cannot necessarily be written in the form
$X_t=\sigma(t) \eta(t)$ where $\eta(t)$ is a cyclostationary process.
Second, even if each $s_c(l)$ is modulated in an identical manner, the corresponding
$\eta(t)$ is not stationary but cyclostationary.

Some care also needs to be used in designing the modulating functions $\{a_c(t)\}$.
The diagonal structure in the Lo\`eve spectrum is preserved as broadened lines
only if $a_c(t)$ is varying more slowly compared to the oscillation $e^{2i\pi t c/D}$.
Otherwise, the diagonal lines will appear as ``shifted'' in frequency.
One option would be to constrain the support of the Fourier transform of $a_c(t)$
so that it is sufficiently well within $-c/D$ and $c/D$. This constraint is often
encountered in signal processing when modeling an amplitude varying oscillation giving
rise to a signal that is strongly related to the types of processes that we wish to
describe, see e.g. the common constraints put on amplitudes in Bedrosian's theorem
\cite{cohen}. In addition to amplitude modulation, it would also be natural in some scenarios to include frequency modulation and replace $a_c(t)$ by
$a_c(t)e^{i\varphi_c(t)}$ in Equation \eqref{cyclostationary2}. Care would have to be taken to keep the average modulation moderate, e.g. $\int a_c^2(t)\varphi_c'(t)dt/
\int a_c^2(t')dt'\approx 0$, as otherwise the process changes characteristics. We investigate this further in the examples in Section \ref{Examples}.

\begin{figure}[t]
  \begin{center}
    \begin{minipage}[]{0.45\textwidth}
      \centering
       (a) \\
      \includegraphics[scale=1]{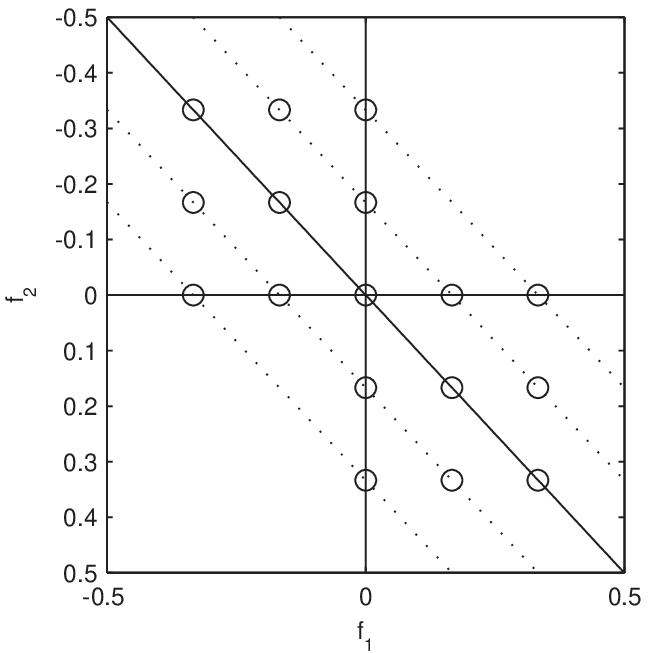}
       \end{minipage}
    \begin{minipage}[]{0.45\textwidth}
      \centering
    (b)\\
  \includegraphics[scale=1]{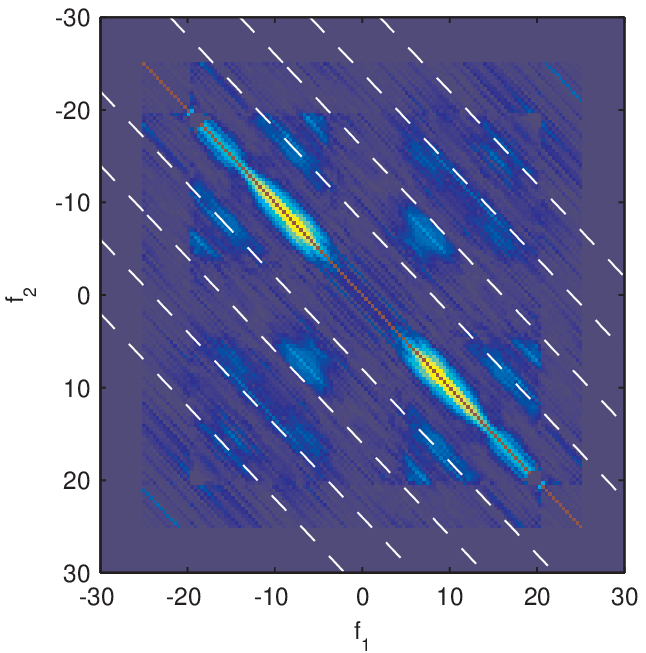}
  \end{minipage}
  \\ 
    \begin{minipage}[]{0.45\textwidth}
      \centering
      (c)
      \\
      \includegraphics[scale=1]{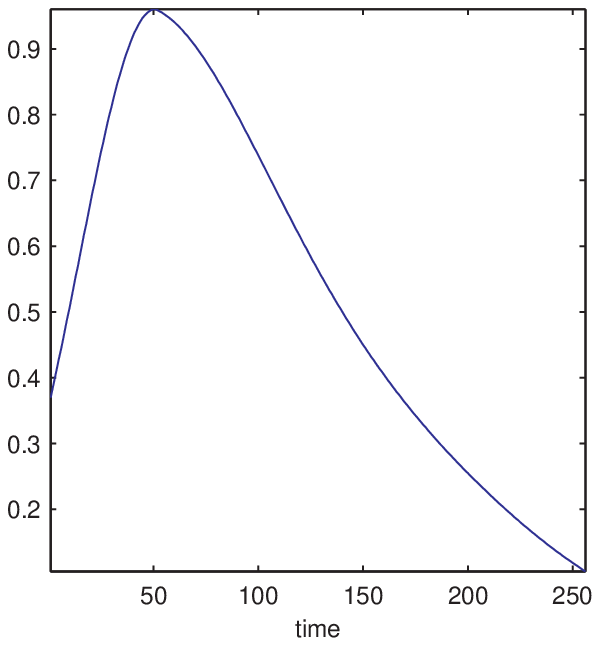}
      \end{minipage}
    \begin{minipage}[]{0.45\textwidth}
      \centering
      (d)\\
     \includegraphics[scale=1]{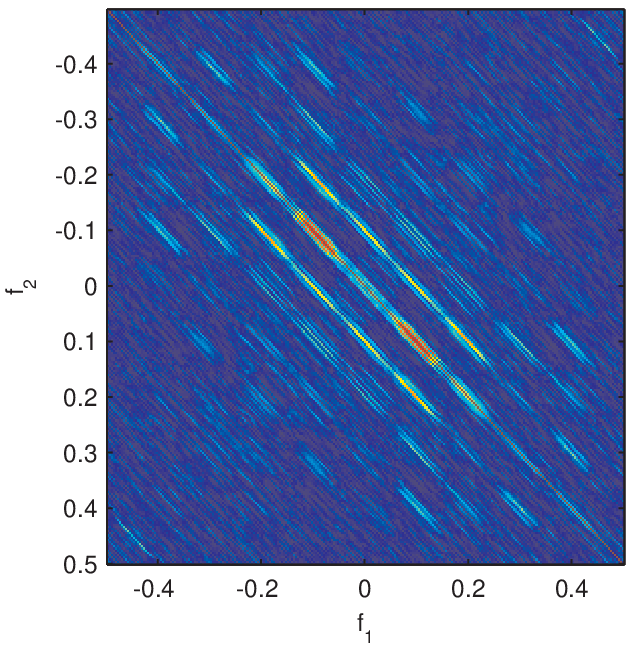}
     \end{minipage}
     \end{center}
  \caption[1]{Plot (a) illustrates that by giving a $\delta$ contribution to the spectrum at frequency $f'=\pm 1/6$ the cyclostationary multiplication leads to a plethora of contributions across the dual-frequency plane. Plot (b) is the average of cross Lo\`eve coherences for the real data over 20 trials. Plot (c) is the within trial modulation function for a simulated example. Plot (d) is the average cross-coherence of a uniformly modulated cyclostationary process over 20 simulated data.}
  \label{fig1}
\end{figure}

We find that the Lo\`eve spectrum is then an aggregation of $2C+1$
components taking the form
\begin{eqnarray}
\label{osci}
S_c(f+\nu,f)&=&\sum_{t,l} \varsigma_c(t,l)e^{2i\pi (t\nu+l f )}
=\sum_{t,l} s_c(l)a_c(t)e^{-2i\pi t c/D}e^{2i\pi (t\nu+l f )}=S_c(f)A_c(\nu-\frac{c}{D}).
\label{osci2}
\end{eqnarray}
Comparing Equation \eqref{cyclostationary} to Equation \eqref{osci2}, we see the difference
between the modulated and the standard cyclostationary model:
the multiplication with the amplitude function $a_c(t)$ in Equation \eqref{osci}
will yield a broadening of lines centered at $\nu=c/D$, with a potential asymmetry
around the line dependent on the form of $A_c(\cdot)$. See Figure \ref{fig4}(c)
for an example of the Lo\`eve spectrum estimated from real data and also a synthetic
example in Figure \ref{fig1}(d) with the associated modulating function $a_c(t)$ in Figure \ref{fig1}(c).
Note that the variance across many near lying oscillations will broaden
the populations of oscillations even further, a feature which is consistent
with our trial averages. From the Heisenberg-Gabor principle (see e.g. \cite[p.~151]{Priestley1988}),
when $a_c(t)$ is more transient in time, then the Lo\`eve spectrum is broader or more dispersed in
frequency. Thus, if a signal displays a very brief temporal response, then the response will be very
diffuse in dual-frequency. In addition, as noted, we could introduce frequency modulation to Eqn \eqref{osci2} by replacing $a_c(t)$ by $a_c(t)e^{-i\varphi_c(t)},$ as long as $|\varphi_c'(t)|$ is constrained not to shift the mean frequency.

To connect our results more generally with amplitude and frequency modulated signals more strongly,
let $U_t$ be a zero mean stationary signal and take $K/2$ to be a positive-valued integer so that
\begin{equation}
X_t=\sum_{k=-K/2}^{K/2} \alpha_k(t)e^{2i\pi kt/D}U_t,
\end{equation}
where $U_t$ is a stationary process with autocovariance sequence $\tilde{s}(l)$. Then
\begin{eqnarray*}
\cov\left\{X_t,X_{t-l} \right\}&=&\sum_{k=-K/2}^{K/2} \sum_{k'=-K/2}^{K/2} \alpha_k(t)
e^{i2\pi kt/D}\alpha_{k'}(t-l)
e^{-i2\pi k'(t-l)/D}\tilde{s}(l)\\
&=&\sum_{c=-K}^K\sum_{v=-K/2+c}^{K/2+c}\alpha_{c+v}(t) \alpha_{v}(t-l)e^{i(2\pi (c+v)t/D-2\pi v(t-l)/D)}\tilde{s}(l).
\end{eqnarray*}
For transparency of expression we investigate a simplified case and let
$
\alpha_k(t)=\alpha_k \alpha(t)$ for $k=-K/2,\dots,K/2$,
and this implies that the form of the covariance takes the simpler form of
\begin{eqnarray}
\nonumber
\cov\left\{X_t,X_{t-l} \right\}
&=&\sum_{c=-K}^{K} \alpha(t)\alpha(t-l)e^{2i\pi ct/D}\sum_{v=-K/2+c}^{K/2+c}
\alpha_{c+v} \alpha_{v}e^{i2\pi
vl/D}\tilde{s}(l).
\end{eqnarray}
To further simplify the covariance we define an additional function $s_c(l)$ and find that
\begin{eqnarray}
\nonumber
\cov\left\{X_t,X_{t-l} \right\}&\approx &\sum_{c=-K}^{K} \alpha^2(t)e^{2i\pi ct/D}s_c(l),\quad s_c(l)=\sum_{v=-K/2+c}^{K/2+c}
\alpha_{c+v} \alpha_{v}e^{i2\pi
vl/D}\tilde{s}(l),
\end{eqnarray}
as long as $\alpha(t)$ is not too variable. Thus we see a simple path to approximate generation of signals with the desired characteristics.

\subsection{Replicated Modulated Cyclostationary Processes}\label{rep-mod}
We have proposed a new class of modulated cyclostationary processes. Now, if we repeatedly observe signals
as realizations of a process, we need to understand the precise notion of replication in the data
generating scheme. For each replicate $r$, we shall assume that the length of the time series is
$T$ and denote the random vector for the $r$-th replicate to be ${\mathbf{X}}^{(r)}=
\begin{pmatrix} X_{1+(r-1)T} & \dots & X_{rT}
\end{pmatrix}$. It is reasonable to assume the probability density function of ${\mathbf{X}}^{(r)}$ to
be from some mixture
\begin{equation}
\label{eqn:mixture}
p\left({\mathbf{X}}^{(r)}\right)=\sum_{m=1}^M \pi^{(rm)}p_{rm}\left({\mathbf{X}}^{(r)}\right),
\end{equation}
that is, the distribution is a mixture between $M$ different states, and where naturally $\sum_m \pi^{(rm)}=1$
for any replicate $r$. As a first step, we shall assume that the $M$ densities
$\{p_{rm}\left(\cdot\right) \}$
 in Equation \eqref{eqn:mixture}
are Gaussian. However, we shall be a bit careful in modeling other aspects of $p_{rm}\left(\cdot \right)$.
In many related problems one would usually take $p_{rm}\left(\cdot \right)=p_{m}\left(\cdot \right)$.
However, this constraint is too stringent and would fail under non-stationarity because time-shifts and phase-shifts
fundamentally change the distribution between replicates, and some $r$-dependence must be inserted back into the model. Here, both a replicate-specific overall time-shift $\delta^{(r)}$ and
between component phase shifts $\phi_c^{(r)}$ will be needed. We shall also permit amplitude modulation per component given by $\theta^{(r)}_c$. We can think of $\delta^{(r)}$ as ``synchronizing'' the {\em whole} waveform observed in ${\mathbf{X}}^{(r)}$, while $\phi_c^{(r)}$ ``synchronizes'' the harmonic components present in ${\mathbf{X}}^{(r)}$, {\em vis-a-vis}
each other.

Then using the newly introduced parameters, we model the common covariance in the replicated $r$ when mixture $m$ only
is present as
\begin{eqnarray}
\label{cyclostationary3}
s^{(rm)}(t,l)&=&\sum_{c=-C_m}^{C_m}\varpi_c^{(rm)}(t,l)=\sum_{c=-C_m}^{C_m} \theta^{(r)}_c a_{c}^{(m)}(t-\delta^{(r)})s_{c}^{(m)}(l)e^{2i\pi
t c/D+i\phi_c^{(r)}},
\end{eqnarray}
where by necessity $\phi_0^{(r)}=0$, $\theta^{(r)}_c\ge 0$ and $\phi_c^{(r)}\in\left(-\pi,\pi\right]$. This model may seem somewhat clumsy but each symbol is directly modulating an aspect of the covariance. With this we find that the covariance in frequency is given by
\begin{eqnarray}
\label{freq-time}
S^{(rm)}(f+\nu,f)&=&\sum_{c=-C_m}^{C_m} S_{c}^{(m)}(f)A_{c}^{(m)}(\nu-\frac{c}{D})\theta^{(r)}_c e^{i\phi_c^{(r)}+2i\pi(\nu-\frac{c}{D})\delta^{(r)} }.
\end{eqnarray}
We therefore determine that the covariance of the increment random process $\{dZ^{(r)}(f)\}$ for the $r$-th replicate
is strongly dependent on the replicate-specific phase-shift $\phi_c^{(r)}$, the time shift $\delta^{(r)}$,
and the amplitude modulation  $\theta^{(r)}_c$. The function $A_c(\cdot)$ will still cause ``blurring'' in frequency and instead
of the $\delta$- function localization of Eqn \eqref{cyclostationary} we
shall observe spread in dual-frequency.
For estimation purposes especially the phase-shift and time-shift in Eqn \eqref{freq-time} are problematic as they cause {\em local} variability in $S(f+\nu,f)$ and so local non-parametric estimators that smooth may perform badly, and the additional dependence of the time and phase shift are unfortunate.
Due to the previously specified locality of $A_{c}^{(m)}(\cdot)$ there will be no substantial overlap in the plane between the different components enumerated by $c$.
We can therefore say that
\begin{eqnarray}
\left|S^{(rm)}(f+\nu,f)\right|&\approx &\sum_{c=-C_m}^{C_m} \left|S_{c}^{(m)}(f)
\right|\left|A_{c}^{(m)}(\nu-\frac{c}{D})\right|\theta^{(r)}_c.
\end{eqnarray}
Note that the variation in magnitude  across replicates is now {\em uniquely} captured by the constant $\theta^{(r)}_c$.
Absolute coherency for the $r$-th replicate using component $m$ near contribution $c$ can then be written as
\begin{eqnarray}
\nonumber
\rho^{1/2}(f_1,f_2)&=&\frac{\left|S^{(rm)}(f_1,f_2)\right|}{\sqrt{\left|S^{(rm)}(f_1,f_1)\left|S^{(rm)}(f_2,f_2)\right|\right|}}
\\
\nonumber
&=&\frac{\left|S_{c}^{(m)}(f_2)\right|\left|A_{c}^{(m)}(f_1-f_2-\frac{c}{D})\right|\theta^{(r)}_c}{
\left|S_{0}^{(m)}(f_1)\right|^{1/2}\left|S_{0}^{(m)}(f_2)\right|^{1/2}\left|A_{0}^{(m)}(-\frac{c}{D})\right|\theta^{(r)}_0}=\rho^{1/2}_{mc}(f_1,f_2)\frac{\theta^{(r)}_c}{\theta^{(r)}_0},
\end{eqnarray}
where $\rho^{1/2}_{mc}(f_1,f_2)$ is modeled in terms of a basis expansion
\begin{eqnarray}
\label{basis}
\rho^{1/2}_{mc}(f_1,f_2)&=&\sum_{p=1}^\infty c_{mc p}\psi_p\left(f_1,f_2\right).
\end{eqnarray}
The basis $\{\psi_p(f_1,f_2)\}$ is chosen iteratively to maximize the magnitude of $c_{mc p}$ for each $p$ starting with $p=1$, and so
Equation \eqref{basis} in fact {\em defines} $\{\psi_p\left(f_1,f_2\right)\}$.
Constructing this representation is generally fine as long as the function $\rho^{1/2}(f_1,f_2)$ is square integrable.
While $\rho^{1/2}_{mc}(f_1,f_2)\ge 0$, this is no longer true for the elements in which we do the expansion, e.g. point-wise $\psi_p(f_1,f_2)$ may be negative, but this should produce no problems. We get that
\begin{eqnarray}
\rho^{1/2}(f_1,f_2)&=&\sum_{c=-C_m}^{C_m} \rho^{1/2}_{mc}(f_1,f_2)\frac{\theta^{(r)}_c}{\theta^{(r)}_0}
=\sum_{c=-C_m}^{C_m}\sum_{p=1}^\infty \frac{\theta^{(r)}_c}{\theta^{(r)}_0} c_{mc p}\psi_p\left(f_1,f_2\right)
=\sum_{p=1}^{\infty} d_{mp}^{(r)}\psi_p\left(f_1,f_2\right).
\label{SVD2}
\end{eqnarray}
We therefore obtain that the coherency $\rho^{1/2}(f_1,f_2)$ for replicate $r$ can be represented in terms of a basis expansion where the weights depend on the trial and the state $m$. As usual $\{\psi_p\left(f_1,f_2\right)\}_p$ are chosen to be orthonormal.

\section{Estimation}\label{Est}

Having developed a non-parametric model that captures the structures we expect to see in the data, we now
develop a procedure for estimating these structures. This is in contrast to statistics, where attention has been focused on very special
forms of nonstationary covariance structures, see e.g. \cite{Lii2002} and \cite{Gardner2006}. Having proposed a general
non-parametric class of models, we now apply a non-parametric estimation technique. Our philosophy is that
by applying non-parametric estimation methods we avoid the bias that is inherently present in parametric models may due to potential omission of features of the data. Ideally once such features have been discovered to be significant they would be put to further scrutiny.
\subsection{Multitaper (MT) Estimation}\label{MT-Est}

Since the covariance structure has been modified by the set of smooth amplitude functions $A_{c}^{(m)}(\cdot)$,
it is reasonable to assume that the Lo\`eve spectrum is smooth. We therefore need a more general
non-parametric estimator, and we turn to \cite{Thomson1982} and the usage of multitaper estimators to this purpose.
We develop a multi-taper procedure for estimating the Lo\`eve coherence from
several time-aligned brain waves recorded from many trials. Denote $X_t^{r}$
to be the time series recorded at the $r$-th trial. Define  $h_t^{(k)}$ to be
the $k$-th orthogonal taper that satisfies $\sum_t [h_{t}^{(k)}]^2 = 1$ and $\sum_t h_{t}^{(k_1)}h_{t}^{(k_2)}=0$ if $k_1\neq k_2$ (for a discussion of tapers and tapering see e.g. \cite{PercivalWalden1993}). The $k$-th
tapered Fourier coefficient at frequency $f$ is then defined to be
\begin{eqnarray*}
x_k^{(r)}(f) & = & \sum_t h_{t}^{(k)} X_t^{r}\exp(-i2\pi f t), \ \ f \in (-\frac{1}{2}, \frac{1}{2}).
\end{eqnarray*}
The $k$-th Lo\`eve periodogram at the frequency pair $(f_1, f_2)$ is defined to be
\begin{eqnarray}
I_k^{(r)}(f_1,f_2) & = & x_k^{(r)}(f_1) x_k^{(r) \ast}(f_2), \ \ (f_1, f_2) \in (-\frac{1}{2},\frac{1}{2}) \times (-\frac{1}{2},\frac{1}{2}) 
\end{eqnarray}
The tapered  Lo\`eve periodogram estimates can be averaged across tapers to produce a suitable
trial-specific non-parametric estimator, the Lo\`eve multitaper spectral estimator,
\begin{equation}
\label{eqn:av:taper}
\overline{I}^{(r)}_\cdot(f_1,f_2)=\frac{1}{K}\sum_k I_k^{(r)}(f_1,f_2).
\end{equation}

Note that the expectation of $I_k^{(r)}(f_1,f_2)$ is not exactly $S(f_1,f_2)$ because of the blurring inherent in using tapers.
It therefore may seem reasonable to weight the sum in \eqref{eqn:av:taper} by the eigenvalues of the localization operator
that produced the tapers $\{h_t^{(k)}\}$, see e.g. \cite{PercivalWalden1993}, but as these are chosen to be so
close to unity in most cases, using a plain average of the estimates with a weighting of unity makes little or
no difference in practice. To obtain an estimator of the ``population" Lo\`eve spectrum, we may take the average
of the trial-specific estimators and can also estimate the coherency in Equation~\eqref{Eq:cohcrosscoh}
\begin{equation}
\label{twice_av}
\overline{I}^{(\cdot)}_\cdot(f_1,f_2)=\frac{1}{R}\sum_r I_\cdot^{(r)}(f_1,f_2),\quad
\widehat{\tau}^{(r)}(f_1,f_2)=\frac{\overline{I}^{(r)}_\cdot(f_1,f_2)}{\sqrt{I^{(r)}_\cdot(f_1,f_1)
I^{(r)}_\cdot(f_2,f_2)}}.
\end{equation}
Using multitaper methods to estimate the Lo\`eve coherency has already been used in Geophysics \cite{Mellors1998}.
%

\vspace{0.25in}

\noindent {\bf Remark.}  We reiterate that the problem with averaging the Lo\`eve periodogram in frequency
and trial lies in the variability of the phase (see Equation \eqref{freq-time}). For a cyclostationary process,
we get a Lo\`eve spectrum that is quite variable in phase across replicates and its phase exhibits smooth variation
along parallel lines. The magnitude in contrast does not depend on replicate-specific time initialization of the cycles.
We believe that over the tapers the estimated phase is stable for a given replicate at a given frequency pair
but that there is variability in where in the cycle we ``start'' across replicates. Averaging phase dependent quantities
in a direction perpendicular to the diagonals across trials therefore makes no sense, even if when we average the real
and imaginary parts we are producing a magnitude weighted average of the phase, which is clearly an improvement.
For this reason Equation \eqref{twice_av} is a ``less than optimal" summary of the population characteristics of the Lo\`eve
coherence matrices.

Note that $\overline{I}^{(\cdot)}_\cdot(f_1,f_2)
\in {\mathbb{C}}$ and its imaginary part is non-zero for most pairs $(f_1, f_2$). For a sufficiently large number of tapers $K$,
it is reasonable to assume that $\Re\{\overline{I}^{(r)}_\cdot(f_1,f_2)\}$ and $\Im\{\overline{I}^{(r)}_\cdot(f_1,f_2)\}$
are nearly jointly Gaussian (see e.g. \cite{Cambanis1994}, and \cite{Brillinger} for stationary processes). This means that the complex variable $\overline{I}^{(r)}_\cdot(f_1,f_2)$
is nearly complex-Gaussian (see, e.g., \cite[p.~39]{Schreier}).
Moreover, when $K$ is sufficiently large and $X_t^{(r)}$ is {\em stationary} then we argue in the Appendix that
for $f_1 \neq f_2$,
\begin{eqnarray}
\label{dist_of_trial}
\widehat{\tau}^{(d)}(f_1,f_2)\overset{d}{\approx} N_C\left(0,\frac{1}{K}\right),\quad
K|\widehat{\tau}^{(d)}(f_1,f_2)|^2\overset{d}{\approx}\frac{1}{2}\chi^2_2.
\end{eqnarray}

\subsection{Multitaper-Singular Value Decomposition (MT-SVD)}\label{MT-SVD}

We have $R$ individual replicate-specific estimates of the square-root coherency $\widehat{\rho}_r^{1/2}(f_1,f_2)$.
Because we need to implement digital processing, we sample the frequency space to
$\{\mathbf{f}_1,\dots,\mathbf{f}_N\}={\mathscr{F}}$, where $
\mathbf{f}_n=(f_1^{(n)},f_2^{(n)})$ is the $n$-th pair of fundamental frequencies. These frequencies do not necessarily
cover the spectrum in the Nyquist range and hence we shall discuss what frequencies to include subsequently. We construct a vector
${\mathbf{p}}^{(r)}=\begin{pmatrix} \widehat{\rho}_r^{1/2}(\mathbf{f}_1) & \dots & \widehat{\rho}_r^{1/2}(\mathbf{f}_N)\end{pmatrix},$
and from the full set of vectors we form
\begin{eqnarray}
{\mathbf{P}}^T=\begin{pmatrix}
{\mathbf{p}}^{(1)T} & ,
\dots &,
{\mathbf{p}}^{(R)T}\end{pmatrix},\quad {\mathbf{P}}={\mathbf{U}}\bm{\Xi}{\mathbf{V}}^H=\sum_k {\xi}_k\mathbf{u}_k \mathbf{v}_k^H.
\label{matriccy}
\end{eqnarray}
We can therefore chose to take
${v}_{kn}=\psi_k\left(\mathbf{f}_n\right),$
and thus expand the matrix in terms of this basis.
We would recognize similar shapes in the dual-frequency domain by examining $\{\mathbf{u}_k\}.$
Now we do not wish to implement this procedure straight on the raw matrix $\widehat{\mathbf{P}}$
as the sampling characteristics of large matrices will make us ``learn noise''. We shall start by determining which part of the matrix is really non-zero.
\subsection{Thresholded Multitaper-Singular Value Decomposition (TMT-SVD)}\label{TMT-SVD}
We first test whether the Lo\`eve spectrum is zero at
off-diagonal points $(f_1,f_2)$ where $f_1 \neq f_2$, as this will help us shrink the
estimates of the coherence for each replicate, and thus remove spurious non-zero coherence.
It would be tempting to compare raw coherency across replicates, but we remind the reader of the incoherent
phase between replicates, as described by Equation \eqref{freq-time}.
We threshold the individual coherency entries by using the distribution of Equation \eqref{dist_of_trial}, implementing a conservative False Discovery Rates correction (FDR) \cite{Efron} to avoid multiple testing issues, but accounting for Hermitian symmetry of the data. This produces
${\mathbf{P}}^{({\mathrm{ht}})}$, from which a singular value decomposition can easily be calculated, resulting in $\mathbf{u}_k^{({\mathrm{ht}})}$ (the trial specific vector), $\xi_k^{({\mathrm{ht}})}$ (the singular values) and $\mathbf{v}_k^{({\mathrm{ht}})}$
(the frequency structure vector). We use $\xi_k^{({\mathrm{ht}})}$
to determine how many wave-forms we need to keep in, in order to describe most of the covariance. We use $\mathbf{u}_k^{({\mathrm{ht}})}$ as a basis from clustering the trials, using the $k$-means method \cite{Gan2007}.

\section{EEG Example}\label{Examples}

\noindent {\bf Overview of the EEG Analysis}.
Brain patterns operate at given ranges of frequencies (see \cite{Sanei}). Most studies concern frequencies corresponding to
the (i.) delta band (0-4 Hertz) which appears in adult slow wave sleep; (ii.) theta band (4-8 Hertz)
which is believed to be associated with the inhibition of elicited responses; (iii.) alpha band
(8-12 Hertz) which is present when eyes are closed and is also associated with inhibition control;
(iv.) beta band (12-30 Hertz) which is present during alert and active states; (v.) gamma band (30-100 Hertz)
which are thought to represent the highly synchronous activity of neurons in response to
a specific cognitive or motor function.
To examine oscillatory properties of single-channel brain signals, scientists use
spectral analysis methods \cite{Priestley1981} which decompose the covariance of an observed time series
according to the independent contributions (weightings or variance) of complex exponentials associated
with different frequencies.  Empirical analyses of our EEG dataset suggest the existence of
coupling or dependence between coefficients at the alpha ($8-12$ Hertz) and beta
($12-30$ Hertz) bands. Such dependence should not be ignored as they could turn out to be important
biomarkers for differentiating between different cognitive processes and mental states.
While this concept of cross-frequency coherence is potentially powerful, it is not commonly
used in practice because of the lack of methods for testing significance of these dependence measures.
In this paper, we elucidate on this concept of spectral dependence and use this to further
interrogate the visual-motor EEG dataset.

\begin{figure}[t]
  \centering
   \begin{minipage}[]{0.45\textwidth}
      \centering
       (a) \\
       \includegraphics[scale=1]{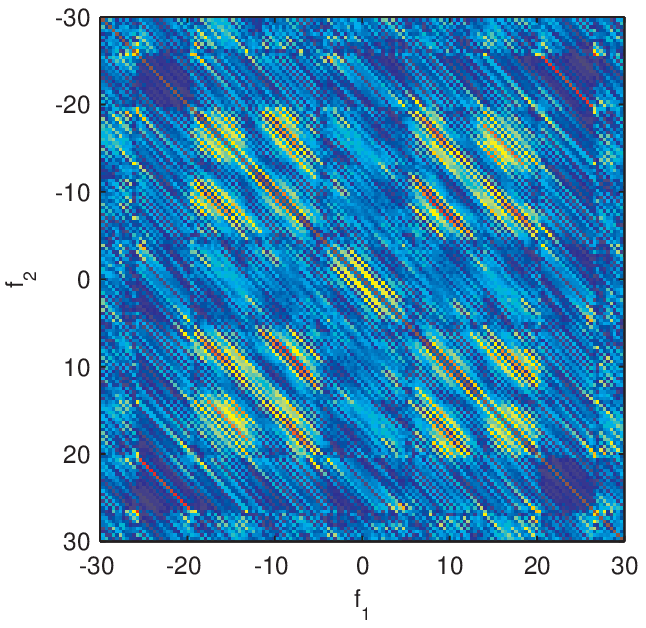}
       \end{minipage}
 \begin{minipage}[]{0.45\textwidth}
      \centering
       (b) \\
       \includegraphics[scale=1]{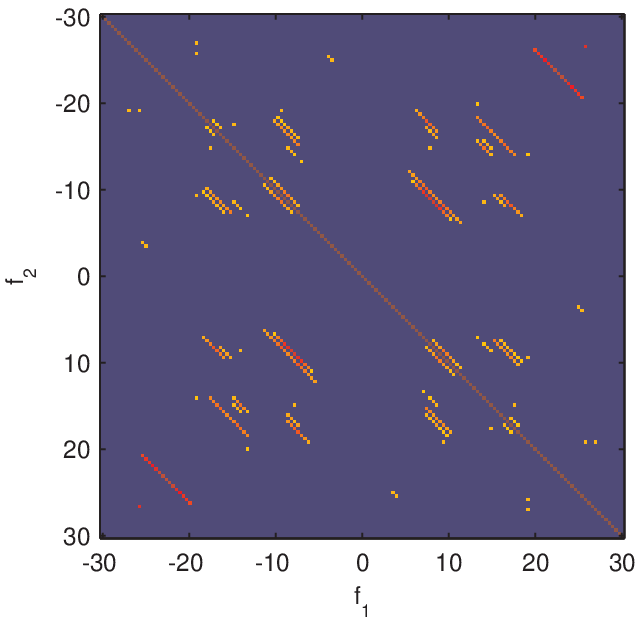}
       \end{minipage}
       \begin{minipage}[]{0.45\textwidth}
      \centering
       (c) \\
        \includegraphics[scale=1]{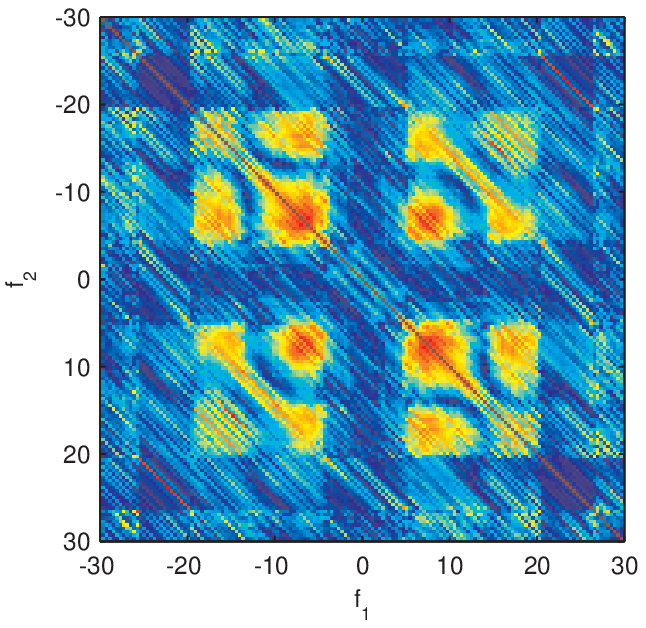}
       \end{minipage}
         \begin{minipage}[]{0.45\textwidth}
      \centering
       (d) \\
    \includegraphics[scale=1]{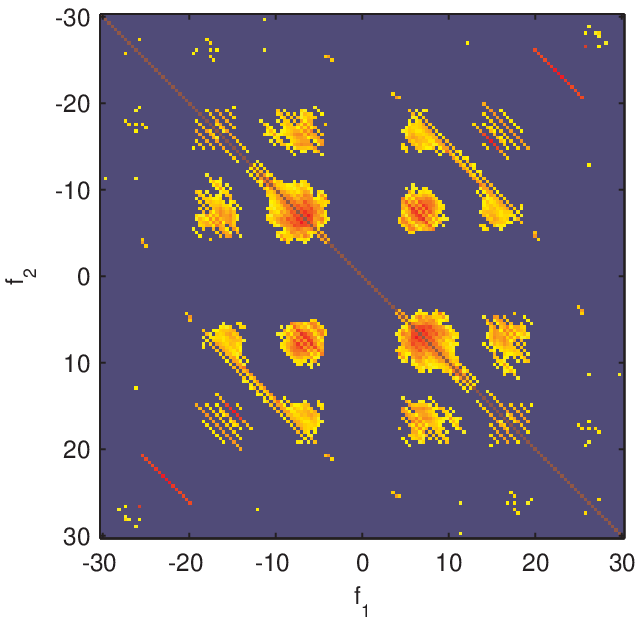}
       \end{minipage}
   \caption[1]{Plots (a) and (c): Two trials of correlations across frequencies
showing two types of structure.  Plots (b) and (d): Corresponding thresholded
versions using FDR (b) and (d). These are trials $b=1$ and $b=8$.}
  \label{fig4}
\end{figure}
\begin{figure}[t]
  \centering
   \begin{minipage}[]{0.45\textwidth}
      \centering
       (a) \\
       \includegraphics[scale=1]{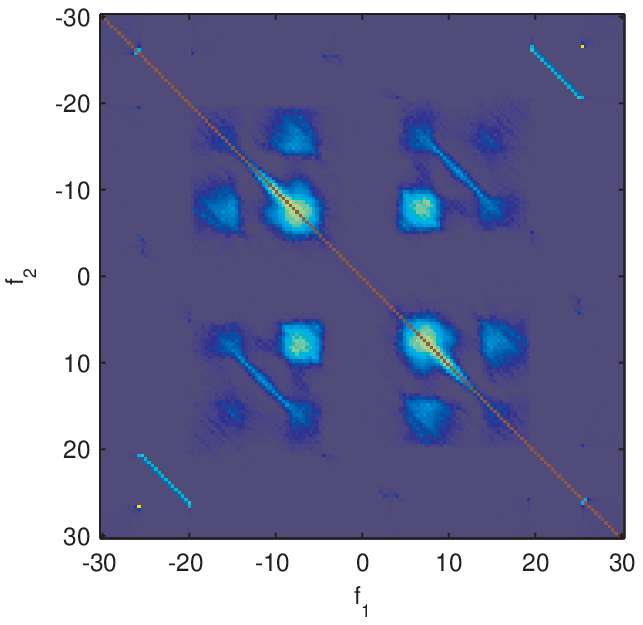}
       \end{minipage}
 \begin{minipage}[]{0.45\textwidth}
      \centering
       (b) \\
       \includegraphics[scale=1]{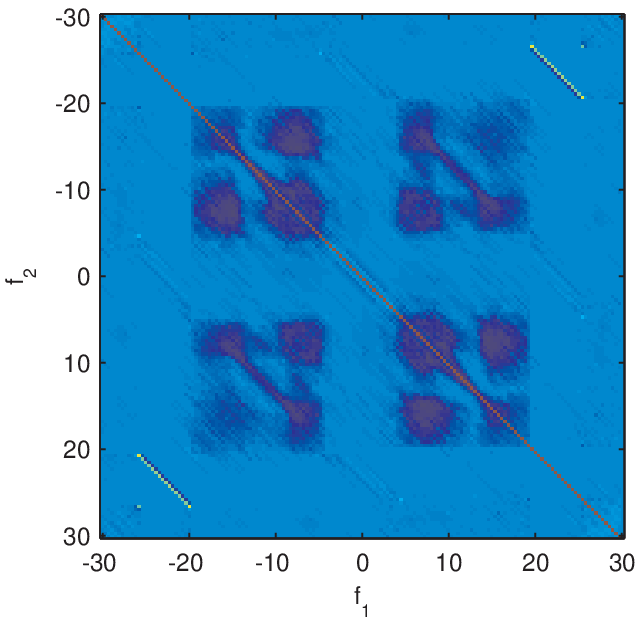}
       \end{minipage}
       \begin{minipage}[]{0.45\textwidth}
      \centering
       (c) \\
       \includegraphics[scale=1]{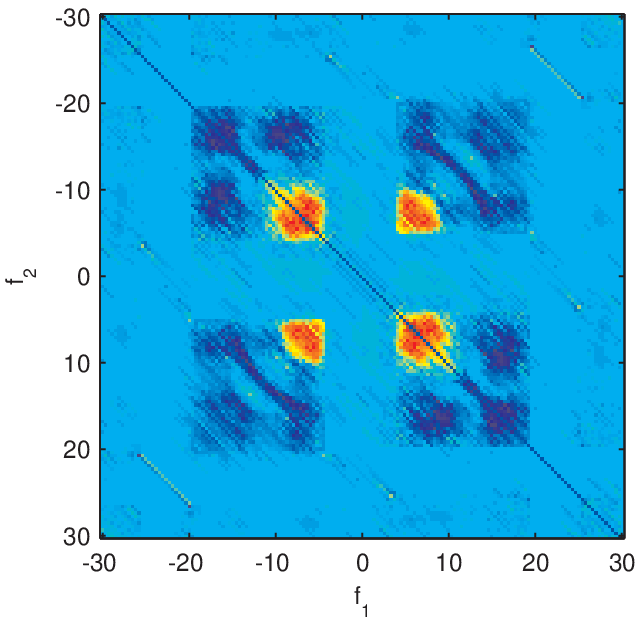}
       \end{minipage}
         \begin{minipage}[]{0.45\textwidth}
      \centering
       (d) \\
       \includegraphics[scale=1]{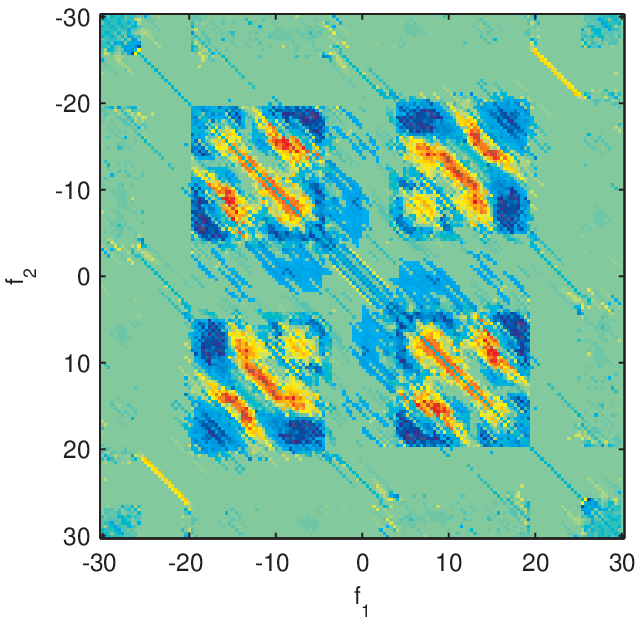}
       \end{minipage}
   \caption[1]{Plots (a) -- (d): The first through fourth principal components of the thresholded
   cross-frequency coherence matrix.}
  \label{fig5}
\end{figure}

\vspace{0.25in}

\noindent {\bf Description of the Data.} The EEG data in this paper is recorded from
a healthy male college student from whom we recorded potentials from the scalp using a
64-channel EEG system (EMS, Biomed, Korneuburg, Germany). The electrodes were applied to the
scalp using conventional methods arrayed in the standard International 10-20 system with
two electrodes that served as a ground and a reference. The EEG was recorded
at 512 Hertz and then band-pass filtered at (0.5,100) Hertz. An additional notch
filter at 60 Hertz was applied to remove the artifact caused by the electrical
power lines. 

The participant performed a visually-guided hand movement
task where he viewed a video monitor, placed about 1 meter away, and responded to
targets that jumped to the left or right from a central position. A target jump,
occurring every 1.5–5 second, instructed the participant to displace the
lever of a hand-held joystick from a central upright position to realign the visual
representation of the joystick orientation with the displaced target,
either to the right or left of center. He received instructions
to start and to move quickly and accurately, and to return the joystick
to the center position only when the target jumped back to the center
of the video monitor. We analyzed EEG signals for 138 rightward movements from
the center position. From the montage of 64 scalp electrodes, we selected the
FC3 electrode, presumably placed over the prefrontal cortex, which is
demonstrated to be implicated in premotor processing \cite{Marconi2001}. The prefrontal cortex, in coordination with the parietal cortex (which is
responsible for visual sensation) and the occipital cortex (which plays a key role
in visual-motor transformations) are all engaged processing and execution of
this visually-guided hand motor task.

\vspace{0.25in}

\noindent {\bf Results and Discussion.} We produced an estimate of $\tau^{(r)}(f_1,f_2)$ for each trial $r$. For illustrative
purposes, in Figure~\ref{fig4}
(a) and (c), we computed $\widehat{\tau}^{(r)}(f_1,f_2)$ for trials $r=1$ and $r=8$.
Of course not all non-zero  Lo\`eve
dual-frequency coherence is really statistically significant. For each trial, we thresholded the
estimated  Lo\`eve dual frequency spectra using the marginal distribution of Equation \eqref{dist_of_trial} combined with the False
Discovery Rates (FDR) procedure with a set rate of $5\%$ \cite{Efron}. To avoid Hermitian redundancy and dependence, the  Lo\`eve dual-frequency coherence matrix was subsampled to half
its size over the frequency interval of interest $[-30,30]$ Hz before this procedure was applied.
Two examples of  estimated  Lo\`eve coherence, derived from the thresholded  Lo\`eve spectra, are
shown in Figure~\ref{fig4} (b) and (d). We see two structures that clearly emerge
from the thresholded spectra: one is the presence of ``thin" lines and the other is the presence of
fat ``blobs". The thin lines can be explained by the structure of a normal cyclostationary model, the fatter structure requires more modulation, like in Example 2.
The blobs are not an effect of the multi-taper method even if this method essentially smooths the
power across frequency: if the smearing was {\em only} due to a resolution issue then the two types of Lo\`eve spectra would not appear in the data. A plausible explanation of the spread is the distribution of oscillations (or
spread of power across some frequency band) and the temporal nonstationarities of the oscillations' amplitudes
within a single trial is discussed in Section \ref{mod-cyc}.

Before combining information across trials,  we first computed the average dual-frequency coherence
across all trials and plotted the average of the magnitudes in Figure~\ref{fig3}, averaged over batches of 20 trials.
One must be cautious when averaging the complex quantities as phase shifts may average real
features to zero. There appears to be some homogeneity of the response across trials, and so the
distribution of magnitude of the coherence averaged over multiple trials is stable and similar
(e.g. Figure \ref{fig3} subplots (a), (b) and (c)). However, these are population results. To see trial-specific 
characteristics, we refer to individual estimates of the Lo\`eve coherence in Figure~\ref{fig4}.

\begin{figure}[t]
  \centering
   \begin{minipage}[]{0.45\textwidth}
      \centering
       (a) \\
       \includegraphics[scale=1]{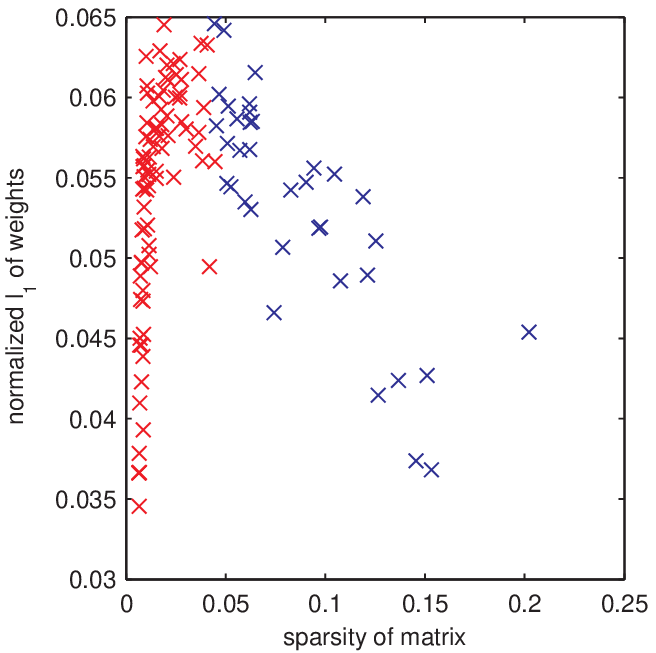}
       \end{minipage}
 \begin{minipage}[]{0.45\textwidth}
      \centering
       (b) \\
       \includegraphics[scale=1]{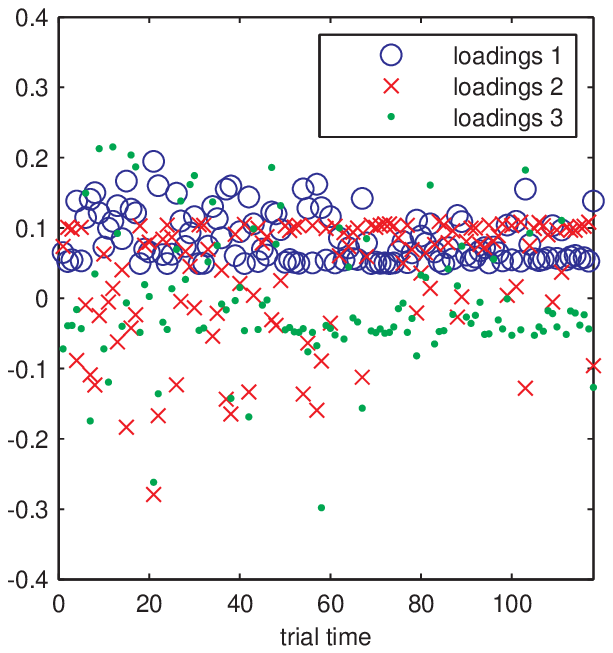}
       \end{minipage}
       \begin{minipage}[]{0.45\textwidth}
      \centering
       (c) \\
       \includegraphics[scale=1]{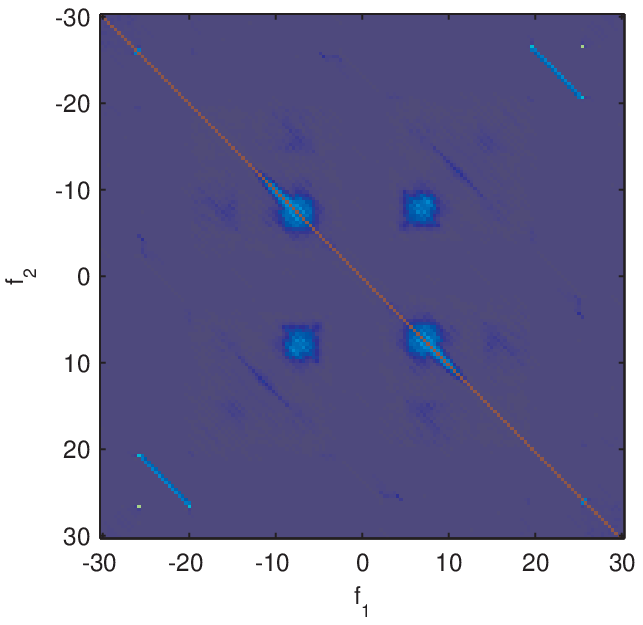}
       \end{minipage}
         \begin{minipage}[]{0.45\textwidth}
      \centering
       (d) \\
       \includegraphics[scale=1]{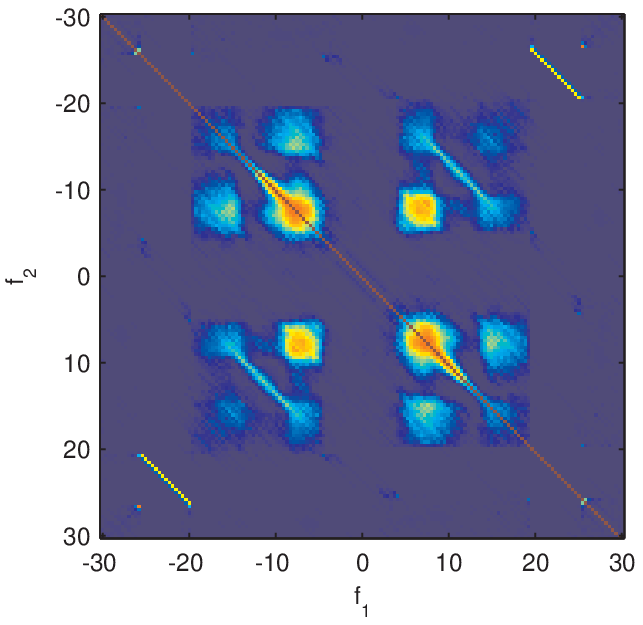}
       \end{minipage}
   \caption[1]{Plot (a): The sparsity of the matrix versus the sparsity across the loadings (recall the weights for each PC is normalized to one). Plot (b): The loads for the first three principal components versus
   trials. Plots (c) and (d): The mean of the two clusters of the weights.}
  \label{fig9}
\end{figure}
A close examination of the individual estimates suggest that there are two
subpopulations to this group of  Lo\`eve (magnitude) coherence matrices. To further investigate this
between-trials structure, we performed a singular value decomposition (or principal component analysis) on the magnitudes of the
thresholded matrices in Figure \ref{fig4}. We plotted the first four principal components in
Figure~\ref{fig5}. The first principal component (PC) is very similar to the averages shown in
Figure~\ref{fig3} (a) and thus the first PC represents a kind of ``average''. The second PC
in Figure \ref{fig5}(b) can either add to or subtract from the ``squares'' of Figure \ref{fig3}(a).
The third and fourth PC in Figure \ref{fig5}(c) and (d) add additional structure.
\begin{figure}[t]
  \centering
   \begin{minipage}[]{0.45\textwidth}
      \centering
       (a) \\
       \includegraphics[scale=1]{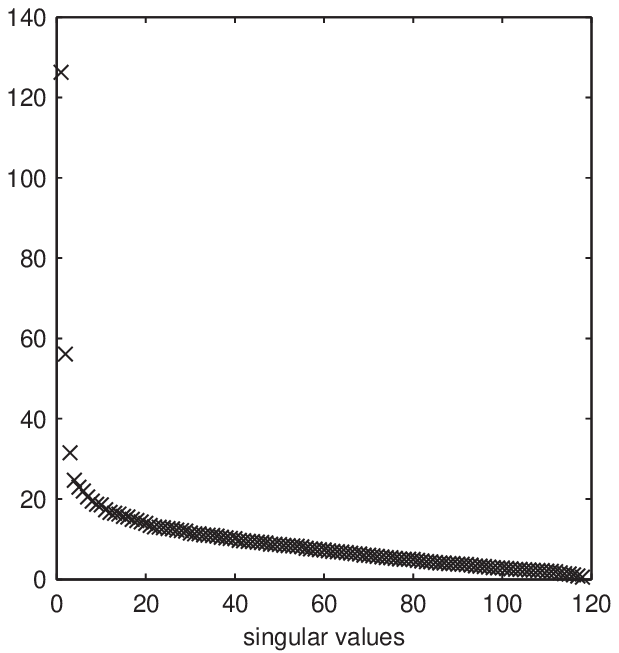}
       \end{minipage}
 \begin{minipage}[]{0.45\textwidth}
      \centering
       (b) \\
       \includegraphics[scale=1]{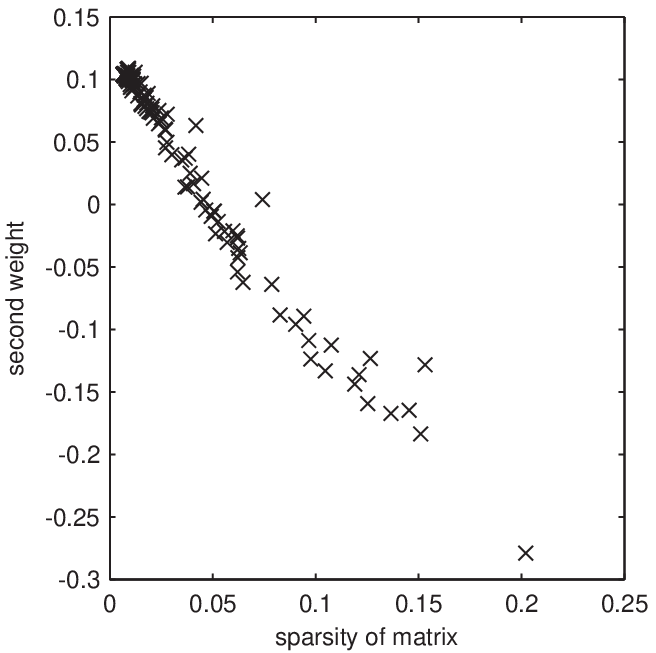}
       \end{minipage}
       \begin{minipage}[]{0.45\textwidth}
      \centering
       (c) \\
       \includegraphics[scale=1]{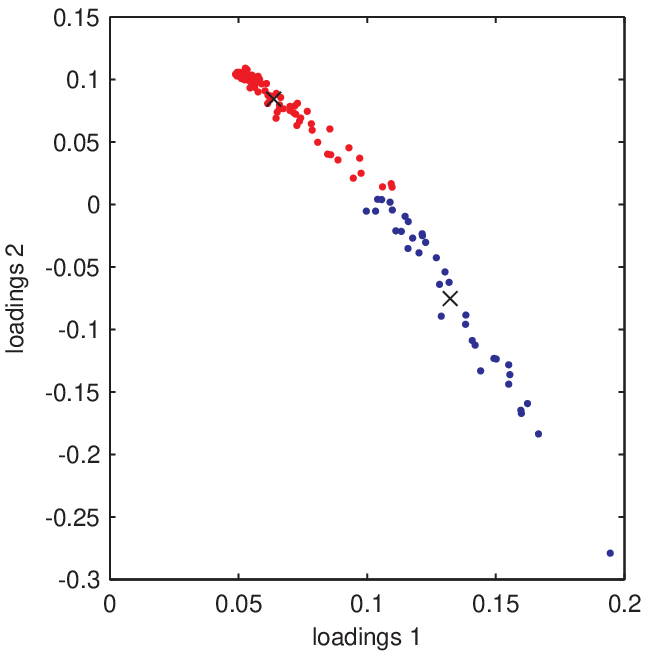}
       \end{minipage}
       \begin{minipage}[]{0.45\textwidth}
      \centering
       (d) \\
       \includegraphics[scale=1]{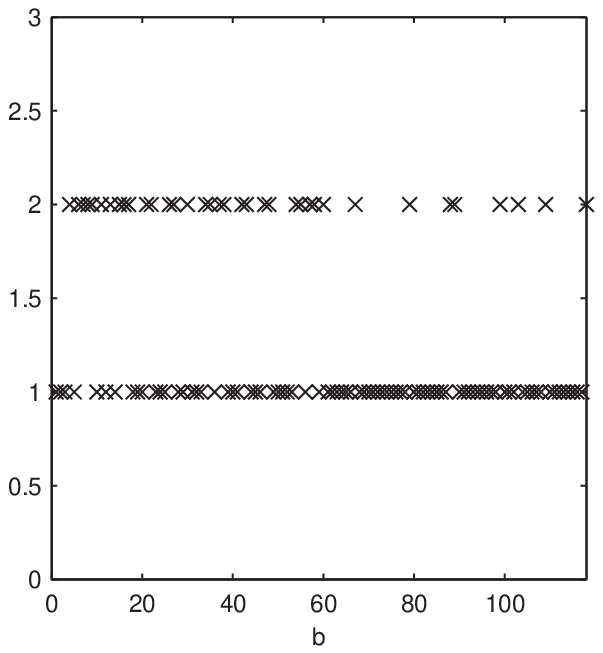}
       \end{minipage}
        \caption[1]{Plot (a): Singular values. Plot (b): The fraction of non-zero values of a matrix
        versus its weight to the second principal component. Plot (c): The first and second weights
        with colors superimposed to signify the clustering from the k-mean method. Plot (d): The
        labelling of each state across trials (d).}
  \label{fig7}
\end{figure}
We plot the singular values in
Figure \ref{fig7}(a). As already noted, the inclusion of the second principal component leads to
substantially sparser correlations and the sparsity of the matrices is clearly correlated
with the second weight (Figure \ref{fig7}(b)). We then used the k-means
method to cluster the trials using only the first two sets of weights. This decision was made
empirically by looking at the weights vs series (e.g. Figure \ref{fig7b}(a) and (b)). The analysis
yielded clearly made up groups, see Figure  \ref{fig7}(c) where the sign of the second loadings
decided the group to which the matrix belongs on trial $b$.
There was a clear trial specific effect and we show the labels of groups versus trials
in Figure  \ref{fig7}(d), where for example cluster two becomes less frequent across trials.

The raw loadings utilized for clustering are displayed in
Figure \ref{fig9}(b). Clearly the longer the subject waited, the higher the likelihood
of diffuse correlation between frequencies, as shown in Figure \ref{fig9}(c) and (d).
The effect of sparsity on the labels was also clear from Figure \ref{fig9}(a).
Red color in the plots corresponds to the matrix in Figure \ref{fig9}(c), which was the
sparser matrix in each trial, but this type of magnitude matrix was less sparse across components for some degrees of
sparsity (see Figure \ref{fig9}(a)). It was less sparse across components because it was not
as ``empty'' as seen from  Figure \ref{fig9}(c), but corresponded to the structure with
some extremely narrow lines as shown in Figure \ref{fig4}(b). This will require several
principal components, each of which will be specific to each frequency that the brain
can produce within a band. The more diffuse structures in Figure \ref{fig4}(d) is the blue
color in the plots and is the fatter matrix in Figure \ref{fig9}(d). These oscillations were
more variable within a trial, as discussed due to variable amplitudes within the trial, causing the narrow lines to spread.

In summary, our proposed modulated cyclostationary model captured highly significant and interesting interactions between the
alpha ($8-12$ Hertz) and beta ($12-30$ Hertz) bands, as clear from Figure \ref{fig9}(d). These results would not have
been obtained from the usual spectral analysis procedures -- which ignore these
types of interactions. Neither does the sparser structure of Figure \ref{fig9}(c) agree with a stationary model as negative and positive frequencies are correlated. This correlation corresponds to a synchrony of oscillations corresponding to starting at a given time after stimulus.

Our findings are interesting because they support the general consensus that the
alpha and beta oscillations
play significant {\it individual} roles during movement. \cite{Pfurtscheller}
demonstrate that beta activity is closely linked to motor behavior and is generally
attenuated during active movements. Moreover, studies have shown beta band activity to be
significant even during just motor imagery and also for task switching. Alpha activity was indicated in
\cite{Moore} to reflect neural activity related to stages of motor response
during a continuous monitoring task. In fact, in a similar response in beta power,
alpha power is reduced at several central electrodes during response execution.
Further studies demonstrated widespread high alpha ($10-12$ Hertz) coherence increase
around the primary sensorimotor cortex during response execution, inhibition and
preparation. While these studies show the contribution of the {\it individual} alpha and
beta activity during movement, our findings suggest the association and some temporal alignment between
these two oscillations which we submit to the neuroscience community for further
interrogation.

\begin{figure}[t]
  \centering
        \begin{minipage}[]{0.45\textwidth}
      \centering
       (a) \\
       \includegraphics[scale=1]{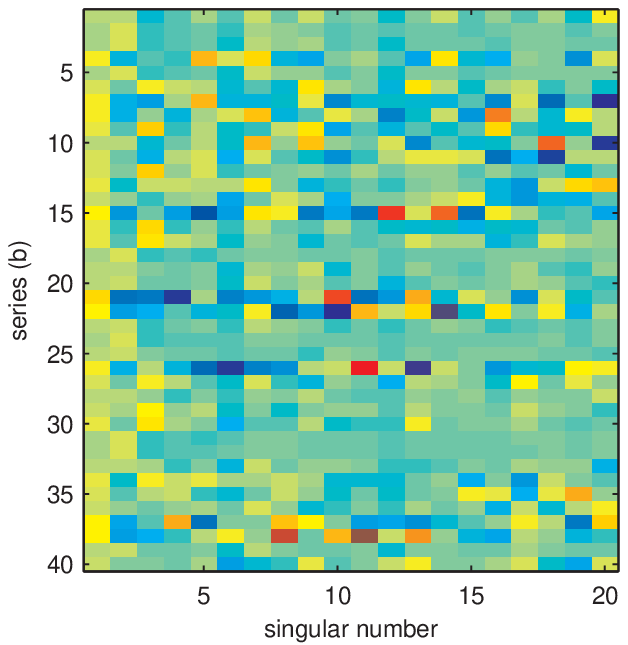}
       \end{minipage}
        \begin{minipage}[]{0.45\textwidth}
      \centering
       (b) \\
       \includegraphics[scale=1]{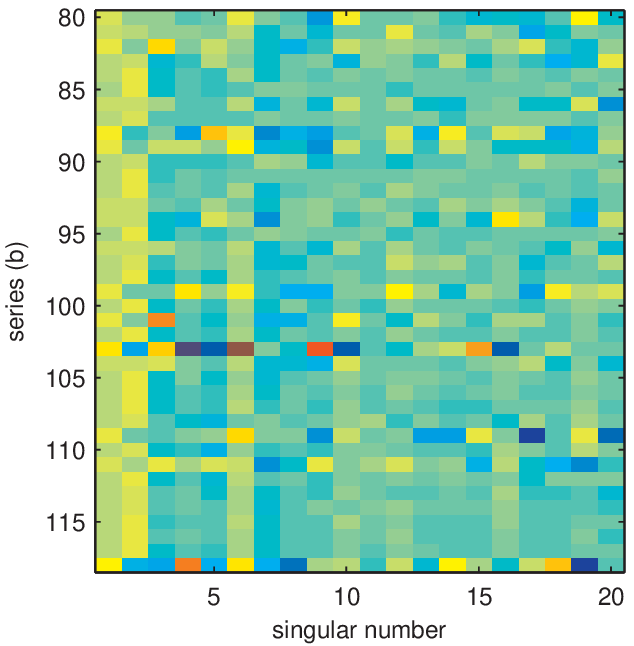}
       \end{minipage}
   \caption[1]{Plot (a): The weights across series and singular vectors for the first 20 trials.
   Plot (b): The weights across series and singular vectors for the last 20 trials.}
  \label{fig7b}
\end{figure}

\section{Discussion}
Stationarity is a traditional assumption that enables consistent estimation of  the covariance of a time series without replicated  measurements. For this reason much of classical estimation theory relies on this assumption. As demonstrated in this paper, there are scenarios when this assumption is unreasonable. However, it is unclear what should be done if the assumption is relaxed, especially if there is no sense of replication between sets of observation.

In the absence of replication, a simpler model must be posited so that ${\cal
O}(T^2)$ covariances need not be estimated. If we insist on retaining a non-parametric
model of covariances, then these must be changing slowly, to enable estimation.
In contrast, with replication, a higher resolution estimate is feasible.
To enable estimation with replication, a model for the form of replication
must be constructed. As we have investigated in this paper, allowing for
any randomness in phase can create destructive interference once combining
information across replicates. We therefore propose to remove the phase of
the Lo\`eve coherence to avoid this destructive interference.

Finally a sensible approach to extracting common features from the replicates
must be designed: we here chose to formulate a model, which prompted us to
use the singular value decomposition. If the entire time series data contains
more than one population, straight averaging is not going to work even if
the phase is removed. With straight averaging, the averages between the two
populations will be recovered instead of a good estimate of each individual
population. We show for simulated and real-life data how the singular value
decomposition can recover the true underlying populations, and help us classify
the state according to their frequency correlation. Smooth and visually appealing
frequency correlation maps are produced from such (see Figures \ref{fig9}(c)
and (d)).

There is an underlying philosophical issue from our analysis. Much effort has focused on defining time series models that are either stationary, or locally stationary in a traditional sense. However real-life data challenge the perception that such is pervasive and the silver bullet for time series applications. Traditional nonstationary models, such as the locally stationary model of Priestley \cite{Priestley1965} and \cite{Dahlhaus2000}, the SLEX (local Fourier) model in \cite{MultiSLEX}, and the locally stationary wavelet model in \cite{Nason2000} are an extremely useful addition to the literature, but cannot capture all features of real data, such as correlation of strongly separated frequencies.  Our addition of the modulated cyclostationary process is to introduce a new model class capable of capturing frequency correlation, but more flexible and parametric model classes are sorely needed. Until such are developed, non-parametric approaches are pivotal to learning additional characteristics of the data. Until such are developed so that we may estimate important features in {\em single trials}, we will need to use repeated trials to highlight and extract important time series characteristics. It is important to always investigate such possibilities when analyzing data, or important aspects of the data will be missed. This will cause us to not infer the correct generating mechanism of the data, and is so very serious.

\appendix

\section*{Statistical Properties of the Fourier Transform}
An arbitrary random
vector $\mathbf{Z}$ which is complex-Gaussian is denoted as $\mathbf{Z}\overset{d}{=}N_C(\bm{\mu},
\bm{\Sigma},\bm{C})$, which means the real and imaginary part are {\em jointly} Gaussian and
\begin{eqnarray}
\Ex\{\mathbf{Z}\}&=&\bm{\mu}_Z,\quad \var\{\mathbf{Z}\}=\Ex\{(\mathbf{Z}-\bm{\mu}_Z)(\mathbf{Z}-\bm{\mu}_Z)^H\}=\bm{\Sigma}_Z\\
\rel\{\mathbf{Z}\}&=&\Ex\{(\mathbf{Z}-\bm{\mu}_Z)(\mathbf{Z}-\bm{\mu}_Z)^T\}=\mathbf{C}_Z.
\end{eqnarray}
In the special case of $\mathbf{C}_Z=\mathbf{0}$ the distribution is complex-Gaussian
proper and is written $\mathbf{Z}\overset{d}{=}N_{C,P}(\bm{\mu}_Z,\bm{\Sigma}_Z)$.
For $\overline{I}^{(r)}_\cdot(f_1,f_2)$, we denote
\[
\bm{\Sigma}_Z = S^{(r)}(f_1,f_2) \ \ {\mbox{and}} \ \ \mathbf{C}_Z = C^{(r)}(f_1,f_2)
\]
By Isserlis' theorem \cite{Isserlis1918},
\begin{eqnarray*}
\Ex \left\{x_k^{(r)}(f_1) x_k^{(r) \ast}(f_2)\right\}&=&S^{(r)}(f_1,f_2),\quad
\Ex \left\{x_k^{(r)}(f_1) x_k^{(r) }(f_2)\right\}=C^{(r)}(f_1,f_2)\\
\var\left\{ x_k^{(r)}(f_1) x_k^{(r) \ast}(f_2)\right\}&=&
S^{(r)}(f_1,f_1)S^{(r)}(f_2,f_2)+\left|C^{(r)}(f_1,f_2) \right|^2\\
\rel\left\{ x_k^{(r)}(f_1) x_k^{(r) \ast}(f_2)\right\}&=&
C^{(r)}(f_1,f_1)C^{(r)\ast}(f_2,f_2)+\left(S^{(r)}(f_1,f_2) \right)^2.
\end{eqnarray*}

For our convenience we define the variance and relation of $\overline{I}^{(r)}_\cdot(f_1,f_2)$ to
be, respectively,
\begin{eqnarray*}
\sigma^2_I(f_1,f_2) & = & \frac{S^{(r)}(f_1,f_1)S^{(r)}(f_2,f_2)+\left|R^{(r)}(f_1,f_2) \right|^2}{K}
\in {\mathbb{R}}^+, \\
c_I(f_1,f_2) & = & \frac{R^{(r)}(f_1,f_1)R^{(r)\ast}(f_2,f_2)+\left(S^{(r)}(f_1,f_2) \right)^2}{K}\in {\mathbb{C}}.
\end{eqnarray*}
Thus, for an arbitrary harmonizable process,
\begin{eqnarray*}
\overline{I}^{(r)}_\cdot(f_1,f_2)&=& {\cal I}^{(r)}_\cdot(f_1,f_2)e^{-i\varphi^{(r)}_\cdot(f_1,f_2)}\quad \overset{d}{=}N_C\left(S^{(r)}(f_1,f_2),\sigma^2_I(f_1,f_2),c_I(f_1,f_2) \right)
\end{eqnarray*}
where ${\cal I}^{(r)}_\cdot(f_1,f_2) \in {\mathbb{R}}^+$.
%
%
If the process $X_t^{(r)}$ is stationary then $S^{(r)}(f_1,f_2)=S^{(r)}(f_1)\delta(f_1-f_2)$ and
thus,
\begin{eqnarray}
\overline{I}^{(r)}_\cdot(f_1,f_2)\overset{d}{\approx}N_C\left(S^{(r)}(f_1)\delta(f_1-f_2),
\frac{S^{(r)}(f_1)S^{(r)}(f_2)}{K},\frac{\left(S^{(r)}(f_1)\right)^2\delta(f_1-f_2) }{K}\right).
\end{eqnarray}
%

\end{document}